\documentclass[referee,a4paper,12pt,traditabstract,dvipdfmx]{swsc} 

\usepackage{amsmath}
\usepackage{graphicx}
\usepackage{txfonts}
\usepackage{subfigure}
\usepackage{lineno}
\usepackage{color}

\DeclareMathOperator*{\argmax}{argmax}

\newcommand{\etal}{et al. }

\begin{document}


   \title{Why do some probabilistic forecasts lack reliability?}
   
   \titlerunning{Why do some probabilistic forecasts lack reliability?}

   \authorrunning{Y\^uki Kubo}

   \author{Y\^uki Kubo}

   \institute{National Institute of Information and Communications Technology,
              Tokyo 184-8795, Japan\\
		\email{kubo@nict.go.jp}
             }


 
  \abstract
   {In this work, we investigate the reliability of the probabilistic binary forecast. We mathematically prove that a necessary, but not sufficient, condition for achieving a reliable probabilistic forecast is maximizing the Peirce skill score ($PSS$) at the threshold probability of the climatological base rate. The condition is confirmed by using artificially synthesized forecast--outcome pair data and previously published probabilistic solar flare forecast models. The condition gives a partial answer as to why some probabilistic forecast system lack reliability, because the system, which does not satisfy the proved condition, can never be reliable. Therefore, the proved condition is very important for the developers of a probabilistic forecast system. The result implies that those who want to develop a reliable probabilistic forecast system must adjust or train the system so as to maximize $PSS$ near the threshold probability of the climatological base rate.
}



   \maketitle

\section{Introduction}
\label{sec:intro}
Forecasts of space weather phenomena have become operational. There are at least two types of forecast of the occurrence of space weather phenomena, namely, deterministic and probabilistic. Because it is difficult to forecast the occurrence of natural phenomena deterministically, a probabilistic forecast is suitable for the occurrence of space weather phenomena, such as solar flare. Moreover, the deterministic forecast can easily be derived by thresholding to a probabilistic forecast (e.g., Jolliffe \& Stephenson 2012). Converting the probabilistic forecast to a deterministic forecast can be performed by forecast users themselves, whose threshold probabilities to determine event occurrence are different. Several authors (e.g., Murphy 1977; Richardson 2000; Zhu \etal 2002) showed in a framework of decision--analytic models that a relative economic value of a probabilistic forecast is higher than that of a deterministic forecast, which meant that a probabilistic forecast is more useful than a deterministic forecast in the sense of economic value. Murphy (1993) mentioned in the sense of forecast consistency that {\it ``Since forecasters' judgments necessarily contain an element of uncertainty, their forecasts must reflect this uncertainty accurately in order to satisfy the basic maxim of forecasting. In general, then, forecasts must be expressed in probabilistic terms."} For these reasons, probabilistic forecast models for the occurrence of space weather phenomena have been developed by several authors. \par

Solar flare occurrence forecasts have been actively studied in the operational space weather forecast community. Recently, many articles related to solar flare occurrence forecasts have been published, which include deterministic forecasts as well as probabilistic forecasts. Examples are human-judged forecasts (Crown 2012; Devos \etal 2014; Kubo \etal 2017; Murray \etal 2017), statistical methods (Wheatland 2005; Falconer \etal 2011; Bloomfield \etal 2012; McCloskey \etal 2016; Steward \etal 2017; Leka \etal 2018), and machine learning forecasts (Bobra \& Couvidat 2015; Muranushi \etal 2015; Huang \etal 2018; Nishizuka \etal 2017, 2018). Many authors have assessed the performance of the forecast models. However, many of the probabilistic forecast models verify a discrimination performance only by using a relative operating characteristic curve, and do not verify other attributes such as reliability, which is one of the most important attributes to be assessed in forecast verification. Several authors (e.g., Jolliffe \& Stephenson 2012; Kubo \etal 2017) mentioned that there are many attributes to be assessed for forecast verification, such as bias, accuracy, discrimination, reliability, and skill. Murphy (1991) pointed out that only one verification measure was not enough to correctly assess the forecast performance due to high dimensionality of a joint probability density of the outcome and forecast. For example, at least three verification measures are required in case of a dichotomous deterministic forecast because the dimensionality in this situation is three. \par

Efforts on comparing the performances of several forecast models have also been in progress. Barnes \etal (2016) compared eleven probabilistic solar flare forecast models and used the relative operating characteristic curve, reliability diagram, and Brier skill score as a verification measures, together with some skill scores for a contingency table created using only one threshold probability of the probabilistic forecast. The reliability diagrams shown in Barnes \etal (2016) showed that several probabilistic solar flare forecast models lack reliability. In the terrestrial weather forecast community, an unreliable probabilistic forecast model is often calibrated (e.g., Gneiting \etal 2007; Primo \etal 2009). However, the calibration of an unreliable probabilistic forecast model is not yet popular in the space weather forecast community. Therefore, it is better that direct outputs from the probabilistic forecast model are already reliable. To realize reliable outputs of probabilistic forecast models, we must investigate the reason why some probabilistic forecast models lack reliability. \par

We investigate a condition for a probabilistic binary forecast to be reliable in this work. In section \ref{sec:mathematics}, we investigate the condition mathematically, and derive a necessary, but not a sufficient, condition for a probabilistic binary forecast to be reliable. In section \ref{sec:confirmation}, the condition will be confirmed by using artificially synthesized forecast probabilities with corresponding outcomes and several probabilistic solar flare forecast models. The discussion and conclusion are described in sections \ref{sec:discussion} and \ref{sec:conclusion}, respectively. \par

\section{Mathematical derivation of the condition}
\label{sec:mathematics}
One of the important attributes to be satisfied for the probabilistic forecast system is reliability. Reliability means a coincidence between the forecast of an event occurrence probability $x$ with the probability density function $p(x) \ (0\le x \le 1)$ and the conditional expectation value of the outcome given the probability $x$; $E(o|x)$ (e.g., Jolliffe \& Stephenson 2012). If a probabilistic forecast system is perfectly reliable, $E(o|x)$ should be equal to $x$. In the case of a binary event, as the outcome $o$ is 1 (100\% probability) for an event and 0 (0\% probability) for no event, $E(o|x)$ can be rewritten as

\begin{equation}
	E(o|x)=\frac{1\cdot p(o =1,x)+ 0\cdot p(o=0,x)}{p(x)}=p(o=1|x),
\label{eq:expectation}
\end{equation}
where $p(o=1,x)$ and $p(o=0,x)$ are joint probability densities of the outcome and forecast. Therefore, the equation

\begin{equation}
	p(o=1|x)=x,
\label{eq:perfect_reliability}
\end{equation}
must be satisfied for a perfectly reliable probabilistic forecast system. \par

By using Bayes' theorem, $p(o=1|x)$ can be rewritten as

\begin{equation}
	p(o=1|x)=\frac{p(o=1) p(x|o=1)}{p(o=1) p(x|o=1) + p(o=0) p(x|o=0)},
\label{eq:Bayes}
\end{equation}
where $p(x|o=1)$ and $p(x|o=0)$ are the conditional probability density functions given the outcome of event and no event, respectively. Hereafter, we refer to $p(x|o=1)$ and $p(x|o=0)$ as $p_1(x)$ and $p_0(x)$, respectively. As $p(o=1)$ is a climatological base rate, we write $p(o=1)$ and $p(o=0)$ as $s$ and $1-s$, respectively. From the equations (\ref{eq:perfect_reliability}) and (\ref{eq:Bayes}), the equation

\begin{equation}
	p_0(x)=\frac{s}{1-s}\frac{1-x}{x}p_1(x)
\label{eq:p0def}
\end{equation}
is derived for a perfectly reliable forecast system. Here, we define the function $f(x)$ as 

\begin{equation}
	f(x) := p_0(x)-p_1(x)=\left(\frac{s}{1-s}\frac{1-x}{x}-1\right) p_1(x).
\end{equation}
$f(x)$ takes zero for $x=s$, positive or zero for $0 \le x < s$, and negative or zero for $s < x \le 1$, because $p_1(x)$ takes a positive or zero value for $0 \le x \le 1$.
\par

Because $p_1(x)$ and $p_0(x)$ are conditional probability density functions given the outcome of event and no event, respectively, the integrals of the functions $p_1(x)$ and $p_0(x)$ from $x$ to $1$ are regarded as a probability of detection ($POD$) and a probability of false detection ($POFD$), respectively, in the forecast verification measure. Therefore, a derivative of the Peirce skill score\footnote{Peirce skill score is also well known as a true skill statistic ($TSS$) and Hanssen-Kuipers ($H-K$) discriminant.} ($PSS=POD-POFD$) by $x$ becomes $f(x)$ As already mentioned, because $f(x)$ takes zero for $x=s$, positive or zero for $0 \le x < s$, and negative or zero for $s < x \le 1$, $PSS(x)$ is maximum at $x=s$. In conclusion, we were able to prove that the proposition,
\begin{equation}
	E(o|x)=x \Rightarrow \argmax_{0 \le x \le 1} PSS(x)=s,
\end{equation}
is true. This means that the maximization of $PSS$ at a threshold probability, which is equal to the climatological base rate, is a necessary condition for a reliable probabilistic forecast.
\par

In the following section, we investigate whether the derived necessary condition is sufficient. If a probabilistic forecast system is unreliable, the conditional expectation value of the outcome given a forecast of the event occurrence probability is not equals to the forecast probability, that is, 
\begin{equation}
	E(o|x)=g(x) \ne x,
\label{eq:not_reliable}
\end{equation}
can be assumed, where $g(x)$ is a function representing a reliability curve. From the equations (\ref{eq:expectation}), (\ref{eq:Bayes}), and (\ref{eq:not_reliable}), the equation

\begin{equation}
	p_0(x)=\frac{s}{1-s}\frac{1-g(x)}{g(x)}p_1(x).
\end{equation}
is derived. As $p_1(x)$ and $p_0(x)$ are conditional probability density functions given the outcome of event and no event, respectively, a derivative of $PSS$ by $x$ is written as

\begin{equation}
	 \frac{dPSS(x)}{dx} = p_0(x)-p_1(x) = \left(\frac{s}{1-s}\frac{1-g(x)}{g(x)}-1\right)p_1(x).
\end{equation}
If there exists a function g(x) satisfying 
\begin{eqnarray}
	\frac{s}{1-s}\frac{1-g(x)}{g(x)}-1=\left\{
	\begin{array}{ll}
	{\rm positive} & (0 \le x < s) \\
	0 & (x=s) \\
	{\rm negative} & (s < x \le 1), \\
	\end{array}
	\right.
\label{eq:gx_condition}
\end{eqnarray} 
then $PSS(x)$ can be maximum at $x=s$, because the derivative of $PSS(x)$ by $x$ takes zero for $x=s$, positive or zero for $0 \le x < s$, and negative or zero for $s < x \le 1$. Actually, because the function
\begin{equation}
	g(x)=s^{1-\beta}x^\beta\ \ (0 < \beta < 1)
\label{eq:gx_function}
\end{equation}
satisfies the equation (\ref{eq:gx_condition}), $PSS(x)$ is maximum at $x=s$ for the unreliable forecast system. Therefore, the proposition,
\begin{equation}
	E(o|x) \ne x \Rightarrow \argmax_{0 \le x \le 1} PSS(x) \ne s,
\end{equation}
is false. This means that the proposition,
\begin{equation}
	\argmax_{0 \le x \le 1} PSS(x) = s \Rightarrow E(o|x) = x,
\end{equation}
is false, and the maximization of $PSS$ at a threshold probability equal to the climatological base rate is a necessary, but not sufficient, condition for a reliable probabilistic forecast system. \par

An important point is that no assumption is made for a functional form of the probability density $p_1(x)$ and $p_0(x)$ when deriving the condition. This means that the condition is independent of the form of the probability density function. \par

\section{Confirmation using forecast data and models}
\label{sec:confirmation}
The necessary condition derived in the previous section is based on continuous probability density functions, which implies that it is based on an infinite number of sample data. However, no infinite number of samples is available in reality. Therefore, the derived condition should be confirmed by using a finite number of sample data. In this section, we confirm the derived condition first by using artificially sampled forecast--outcome pairs and then by using several probabilistic solar flare forecast models described by Barnes \etal (2016). \par

\subsection{Synthetic forecast data}
\label{sec:forecast data}
A probabilistic binary forecast system is fully determined by defining the climatological base rate $s$ and two conditional probability density functions of event occurrence probability, $p_1(x)$ and $p_0(x)$. Synthetic forecast--outcome pairs are randomly sampled from $p_1(x)$ and $p_0(x)$, so as to climatological base rate being $s$. In this article, the climatological base rate $s$ is fixed at 0.1, which represents a somewhat rare event case. The total number of sampled forecast--outcome pairs is 10,000. \par

Because the independent variable of the conditional probability density functions is the probability $x$, the range of $x$ must be from 0 to 1. Therefore, the beta distribution $Be(x; a, b)$ is employed for probability density functions, whose definition appears in Appendix. While a beta distribution can flexibly change its shape depending on the two parameters, it is suitable for investigating various types of situations. Three cases are investigated: (1) perfectly reliable, (2) $PSS$ is maximum at the probability largely different from the climatological base rate, and (3) $PSS$ is maximum at the probability equal to the climatological base rate but unreliable. Although only specific forms of probability density function are considered in the subsequent three subsections, the results of the studies are independent of the form of the probability density function. \par

\subsection{Case 1: Perfectly reliable forecast}
\label{sec:case1}
In case 1, the two conditional probability density functions of event occurrence probability, $p_1(x)$ and $p_0(x)$, are set as $Be(x; 1.1, 0.9)$ and $\left[10Be(x; 0.1, 0.9)-Be(x; 1.1, 0.9)\right]/9$, respectively, so that the two density functions satisfy the equation (\ref{eq:p0def}), which states that the probabilistic forecast system is reliable. Randomly sampled variates from the probability density functions are pooled as the artificial forecast--outcome pairs. \par

\begin{figure}
	\centering
	\subfigure[Reliability diagram for case 1]{
	\includegraphics[width=8cm]{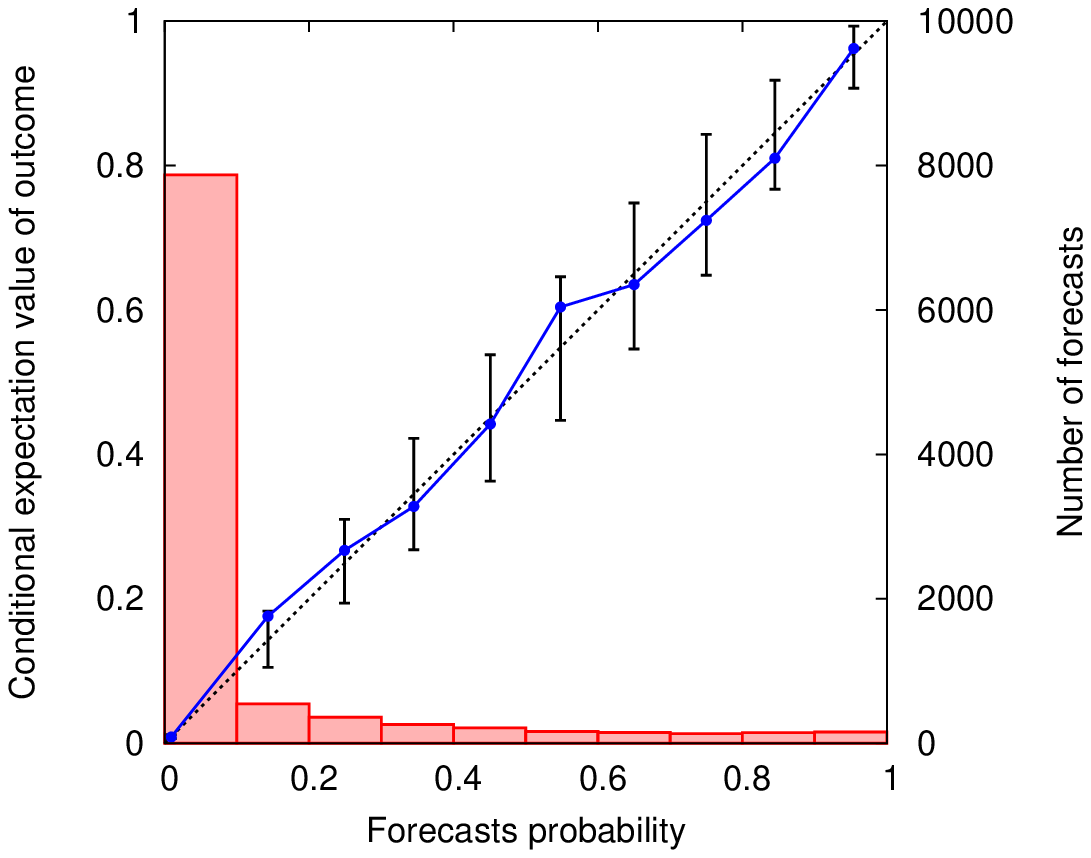}
	\label{fig:reliability_case1}
	}
	\subfigure[$PSS$ vs threshold probability for case 1]{
	\includegraphics[width=8cm]{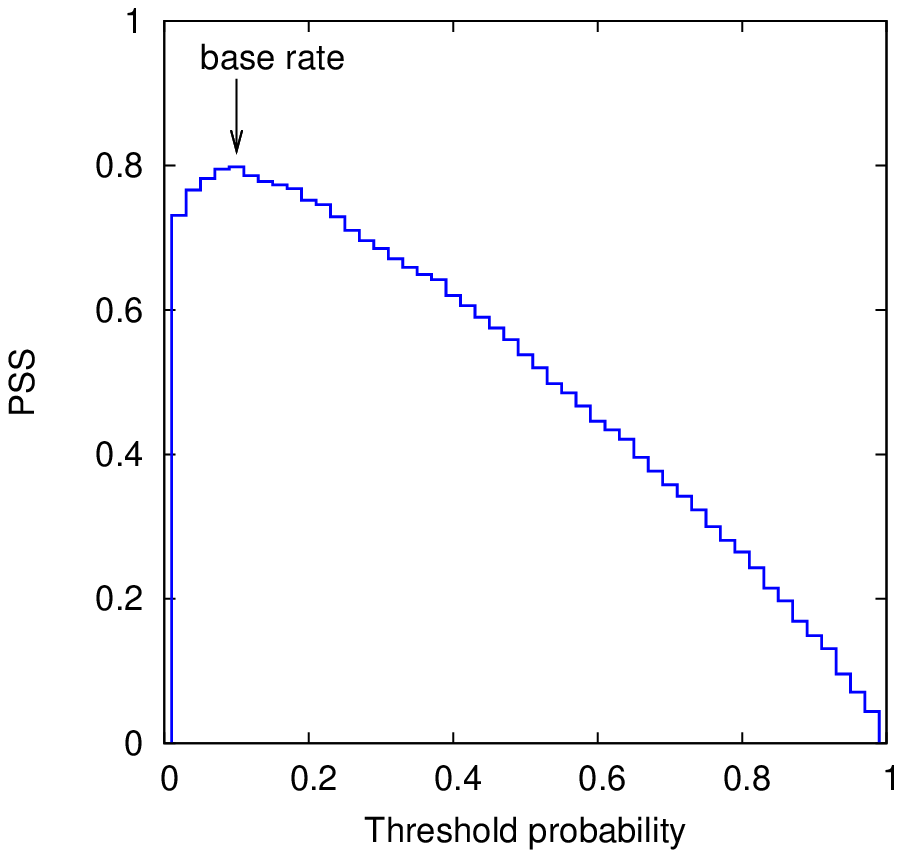}
	\label{fig:PSS_case1}
	}
	\caption{(a) Reliability diagram for case 1. Blue dots connected by lines depict the conditional expectation values of outcome. A perfect reliability curve is depicted by the diagonal dashed line, on which the 99 \% consistency bars are drawn as vertical dashes. Red histograms with the right axis show the number of probabilistic forecasts within bins. (b) $PSS$ versus threshold probability for case 1.}
\label{fig:case1}
\end{figure}

Figure \ref{fig:reliability_case1} shows a reliability diagram for case 1. The blue dots connected by lines depict the conditional expectation values of the outcome. A perfect reliability curve is depicted by the diagonal dashed line, on which the 99 \% consistency bars (Br\"ocker \& Smith 2007, Jolliffe \& Stephenson 2012) are drawn as vertical dashes. The 99 \% consistency bar shows the range within which 99 \% of the conditional expectation value of the outcome given the probability would fall, if it were assumed that the original data is sampled from the perfectly reliable probabilistic forecast system. The red histograms with the right axis show a number of probabilistic forecasts within bins. It is clear that all the conditional expectation values of the outcome are located within the 99 \% consistency bars. This means that the synthetic probabilistic forecast is almost perfectly reliable (of course, it is by definition). \par

According to the condition derived in Section \ref{sec:mathematics}, $PSS$ must be maximum at the threshold probability of 0.1, which is a climatological base rate. Figure \ref{fig:PSS_case1} shows the variation of $PSS$ versus the various threshold probabilities calculated using the synthetic forecast--outcome pairs. We can clearly see that $PSS$ is maximum at around the climatological base rate. \par

\subsection{Case 2: Maximize $PSS$ at a probability different from the climatological base rate}
\label{sec:case2}
In case 2, $p_1(x)$ and $p_0(x)$ are set as $Be(x; 2.2, 0.4)$ and $\left[10Be(x; 0.2, 0.4)-Be(x; 2.2, 0.4)\right]/9$, respectively, for which $PSS$ is maximum at the threshold probability of 0.5, which is largely different from the climatological base rate. \par

\begin{figure}
	\centering
	\subfigure[Reliability diagram for case 2]{
	\includegraphics[width=8cm]{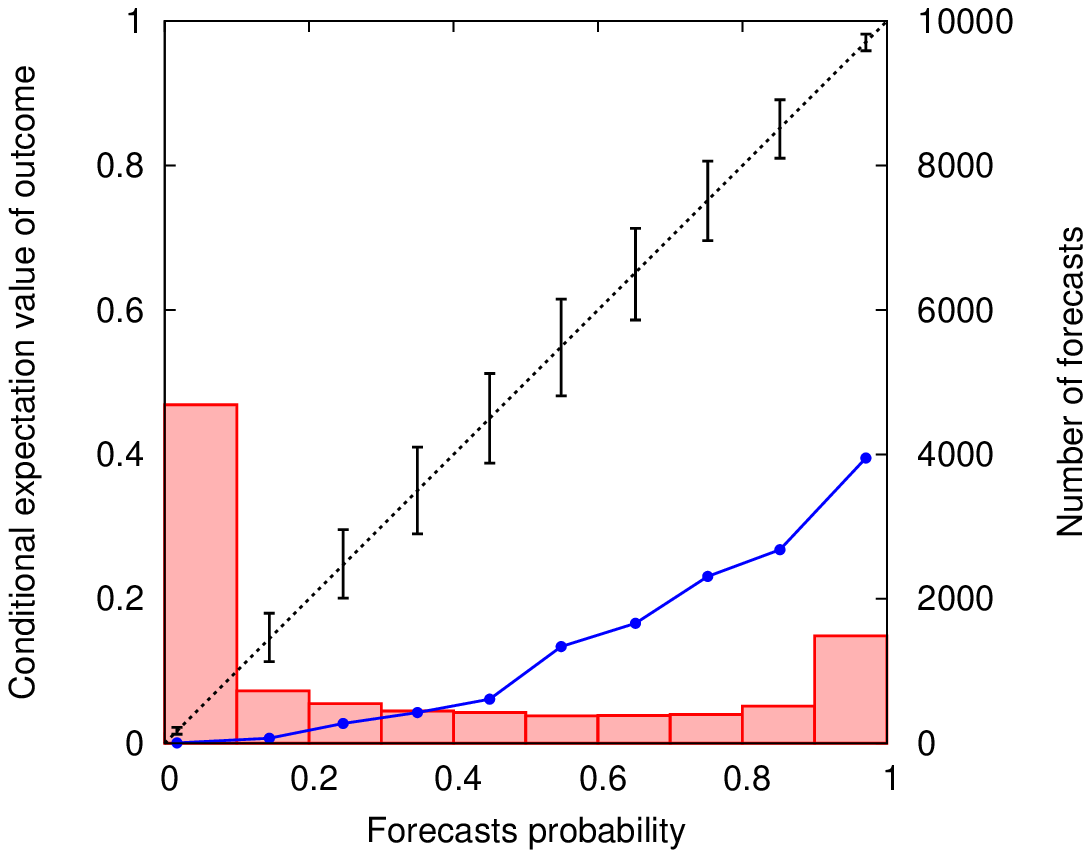}
	\label{fig:reliability_case2}
	}
	\subfigure[$PSS$ vs threshold probability for case 2]{
	\includegraphics[width=8cm]{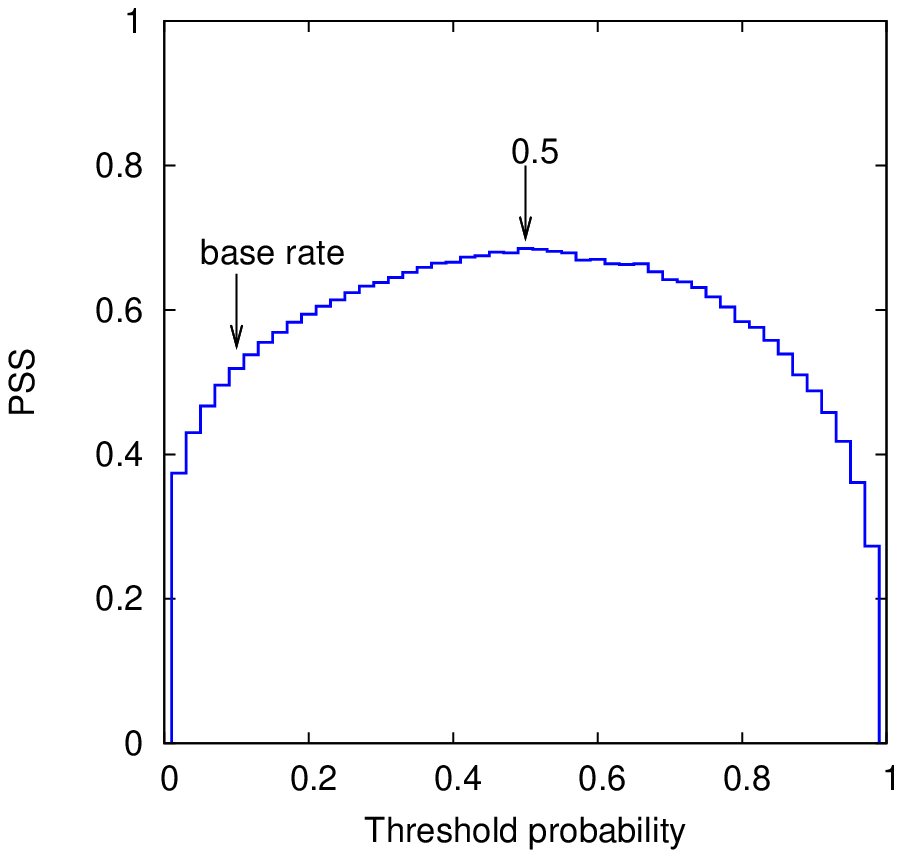}
	\label{fig:PSS_case2}
	}
	\caption{Same as figure \ref{fig:case1}, but for case 2}
\label{fig:case2}
\end{figure}

Figure \ref{fig:PSS_case2} depicts the plot of $PSS$ versus various threshold probabilities. The diagram shows that $PSS$ is maximum at the threshold probability of around 0.5 (by definition), which is far from the climatological base rate. \par

According to the condition mathematically derived in Section \ref{sec:mathematics}, the probabilistic forecast on case 2 must be unreliable. In the following, we will confirm that the forecast is unreliable by drawing the reliability diagram. Figure \ref{fig:reliability_case2} shows a reliability diagram for case 2. The dots, lines, dashes, and histogram represent the same quantities as those in case 1. We can recognize from the figure that the conditional expectation values of the outcome are not on the perfect reliability line. This fact confirms that case 2 is an unreliable probabilistic forecast system. \par

\subsection{Case 3: Maximize $PSS$ at the climatological base rate but unreliable forecast}
\label{sec:case3}
In case 3, $p_1(x)$ and $p_0(x)$ are set as $Be(x; 0.83, 1.19)$ and $\left[10Be(x; 0.23, 1.19)-Be(x; 0.83, 1.19)\right]/9$, respectively, for which $PSS$ is maximum at the threshold probability of the climatological base rate. \par

\begin{figure}
	\centering
	\subfigure[Reliability diagram for case 3]{
	\includegraphics[width=8cm]{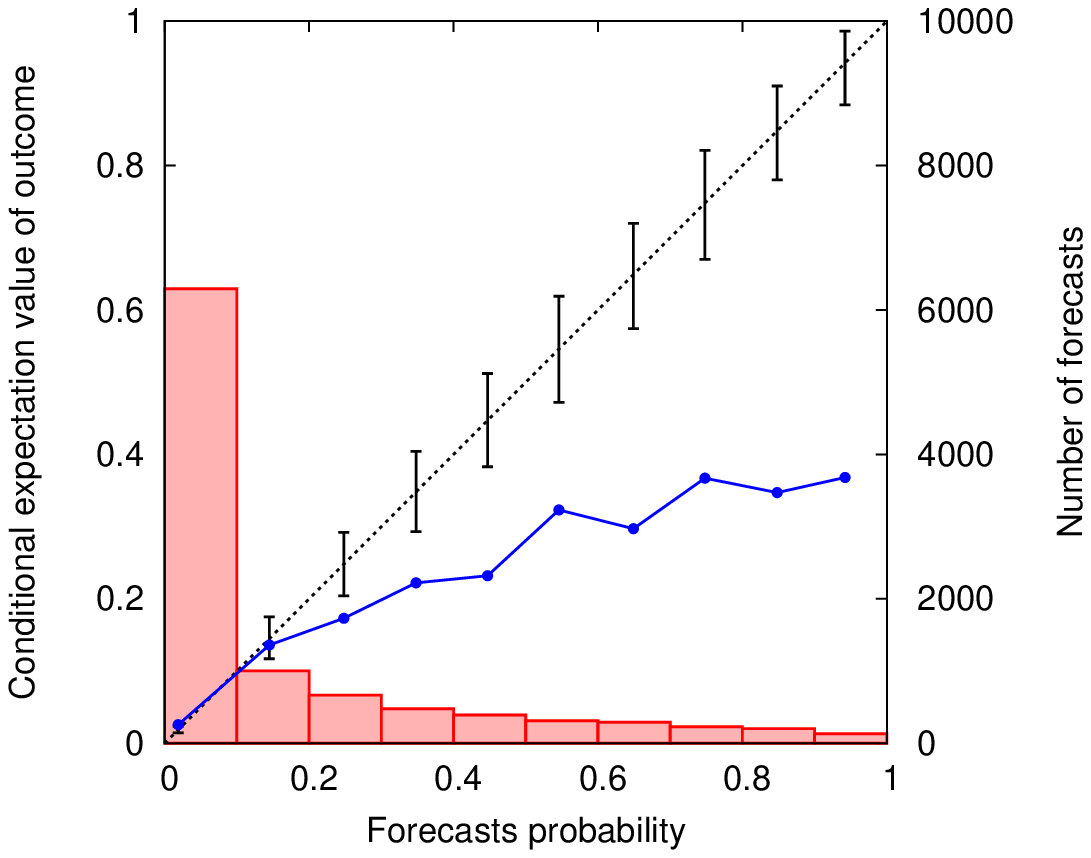}
	\label{fig:reliability_case3}
	}
	\subfigure[$PSS$ vs threshold probability for case 3]{
	\includegraphics[width=8cm]{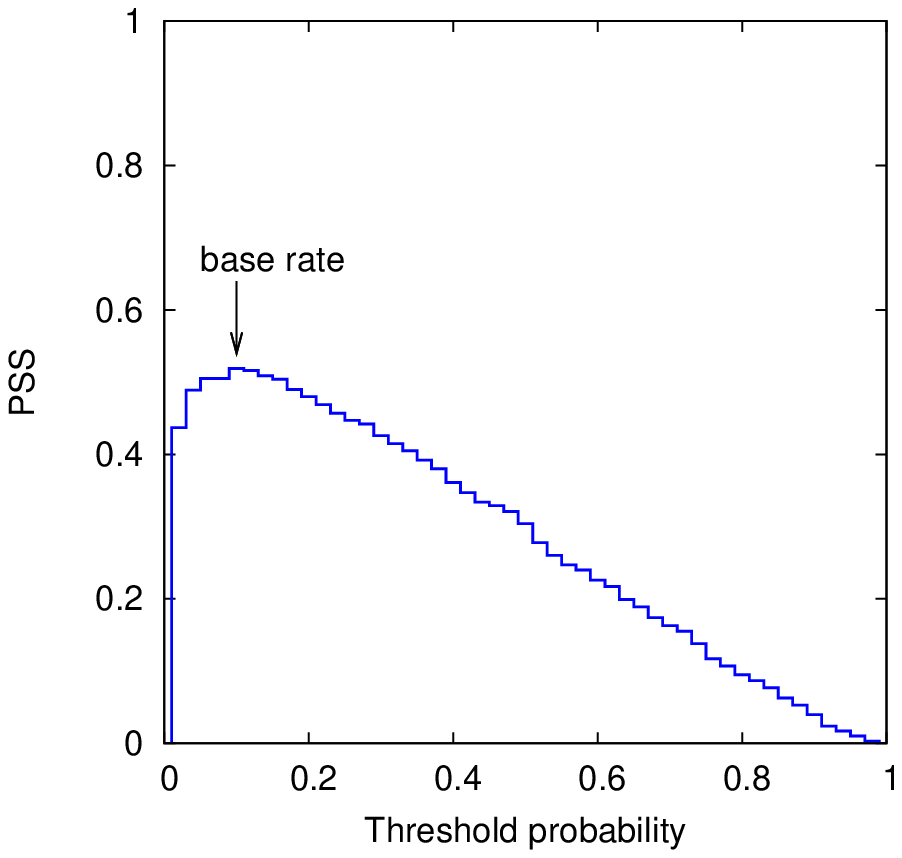}
	\label{fig:PSS_case3}
	}
	\caption{Same as figure \ref{fig:case1}, but for case 3}
\label{fig:case3}
\end{figure}

A plot of $PSS$ versus various threshold probabilities is depicted in Figure \ref{fig:PSS_case3} using a finite number of synthetic forecast--outcome pairs. The figure shows that $PSS$ is maximum at around the climatological base rate (by definition). However, as proven in the previous section, because the maximization of $PSS$ at the threshold probability of the climatological base rate is not sufficient condition for probabilistic forecast to be reliable, whether the probabilistic forecast is reliable should not be decided. To confirm this theoretically derived result, a reliability diagram for case 3 is drawn in Figure \ref{fig:reliability_case3}. The dots, lines, dashes, and histogram represent the same quantities as those in case 1. Clearly, the conditional expectation values of the outcome do not follow a perfect reliability line. This fact shows that the probabilistic forecast is unreliable even if $PSS$ is maximum at around the climatological base rate. \par

\subsection{Solar flare forecast models}
\label{sec:solar_flare_models}
As Barnes \etal (2016) plotted reliability diagrams and estimated threshold probabilities maximizing $PSS$\footnote{$PSS$ was referred as $H\&KSS$ in Barnes \etal (2016).} for eleven solar flare forecast models, these results are used for confirming the validity of the condition derived in this study. Although they dealt with three event definitions, we refer to only one event definition (C1.0 or greater flare) because, as there were few flare event samples for other the two event definitions, the error bars for the reliability diagrams were large. In this subsection, the terms ``table" and ``figure" denote the table and figure that appeared in Barnes \etal (2016) unless explicitly stated. 
\par

Ten models out of eleven can forecast the events of C1.0 or greater flare, and were assessed for the events (figures 11, 12, 13, 17, 19, 20, 22, 23, 25, and 26). Climatological base rates for the ten models were shown in the tables 8, 9, 10, 12, 13, 14, 15, 16, 17, and 18, respectively. Reliability diagrams (top panels) for figures 12, 13, 15, 19, 20, and 22 show that the reliabilities for these models were relatively good (of course, no models has perfect reliability). According to the condition derived in section \ref{sec:mathematics}, the threshold probabilities maximizing $PSS$ for these models should be near the climatological base rate. As the threshold probabilities maximizing $PSS$ are shown in the bottom panels of the figures, we refer to these values. The absolute values of difference between the climatological base rate and the threshold probability maximizing $PSS$ for the relatively reliable forecast models were between 0.015 to 0.049, which shows that the threshold probabilities maximizing $PSS$ were very close to the climatological base rates. On the other hand, the absolute values of difference between the climatological base rate and the threshold probability maximizing $PSS$ for the models shown in figures 11, 23, 25, and 26 were 0.193, 0.268, 0.393, and 0.150, respectively, which meant that the threshold probabilities maximizing $PSS$ were largely far from the climatological base rates. We clearly recognize from the figures 11, 23, 25, and 26 that the reliabilities for these models were relatively poor. This result shows that the model that has a threshold probability maximizing $PSS$ far from a climatological base rate lacks reliability. These results are consistent with the mathematically derived condition. These results are summarized in table \ref{tbl:flare_model} in this paper. \par

From the examples shown in this section, it is confirmed that the maximization of $PSS$ at a climatological base rate is a necessary, but not sufficient, condition for a reliable probabilistic forecast. We used beta distributions to describe the probability densities in the examples. However, we emphasize again that the confirmed result does not depend on the form of the probability density as shown in section \ref{sec:solar_flare_models}, so the result is quite general. \par

\section{Discussion}
\label{sec:discussion}
The condition that $PSS$ is maximum at a threshold probability of a climatological base rate is a necessary condition for a probabilistic forecast system to be reliable. That is, if the probabilistic forecast system is reliable, the $PSS$ of the system is maximum at the threshold probability of, definitely, the climatological base rate. In other words, a probabilistic forecast system whose $PSS$ is maximum at a largely different climatological base rate can never become reliable. This claim is very important for developers of probabilistic forecast systems. Those who want to develop a reliable probabilistic forecast system must adjust or train their system so that $PSS$ is maximum at the threshold probability of the climatological base rate. Of course, the adjustment or training alone is not necessarily enough for a reliable system, because the condition is not a sufficient condition. However, if no adjustment or training is carried out, their system can never become reliable. \par

A joint probability density of forecast--outcome pairs can be factored into a conditional probability density and marginal probability density. In a distribution-oriented forecast verification framework (Murphy \& Winkler 1987), two types of factorization are possible. One is a calibration-refinement factorization, which is a factorization into the conditional probability density of observation given forecast (calibration distribution) and the marginal probability density of a forecast (refinement distribution). The other is a likelihood-base rate factorization, which is a factorization into the conditional probability density of forecast given observation (likelihood distribution) and a marginal probability density of observation (base rate distribution). While an attribute of reliability is directly related with the calibration distribution, $PSS$ is only related with the likelihood distribution, which implies that $PSS$ can say nothing on reliability. That is, the completely different aspects of joint probability density are assessed on the basis of reliability and $PSS$. It is interesting that, despite this fact, a reliable probabilistic forecast is directly related with the maximization of $PSS$. The interesting question as to why the maximization of $PSS$ at the threshold probability of a climatological base rate is related with the reliable probabilistic forecast system, can partly be accounted for by considering the factorization of the joint probability density. A combination of likelihood and base rate distributions can completely describe the joint probability density of forecast--outcome. This means that although the likelihood distribution alone cannot assess a calibration distribution, the combination of likelihood and base rate distributions can do so. Therefore, the combination of information of $PSS$ and the climatological base rate is required for assessing information of reliability. \par

Some related literatures with this study have published in meteorological forecast verification. Richardson (2000) discussed a relative economic value of forecasts in the framework of a decision--analytic models. He mentioned that the maximum relative economic value for a deterministic forecast was attained at the point where an user's cost--loss ratio equals to a climatological base rate and was given by $PSS$\footnote{$PSS$ was referred as $KS$ in Richardson (2000).}. This meant that a maximum relative economic value for probabilistic forecast was given by a maximum $PSS$ under the condition that an user's cost--loss ratio equals to a climatological base rate. The fact that the relationship between a climatological base rate and maximum $PSS$ appears in several kinds of situation for forecast verification is interesting. This point should be further investigated. \par

\section{Conclusion}
\label{sec:conclusion}
We mathematically derived a necessary condition for a probabilistic binary forecast to be reliable. The condition was maximizing a $PSS$ at a threshold probability of a climatological base rate. The condition was confirmed by using artificially synthesized forecast--outcome pair data and several published probabilistic solar flare forecast models. An important point is that the condition is derived without assuming the form of the probability density function. This means that the condition generally holds. This condition is quite important for the developers of probabilistic forecast systems. When a reliable probabilistic binary forecast system is developed, the developer must adjust or train the system so as to maximize $PSS$ at the threshold probability of the climatological base rate. The condition gives a partial answer as to why some probabilistic forecast systems lack reliability because the system that does not satisfy the condition can never be reliable. \par

\begin{table}
	\begin{tabular}{l|rrrrrrrrrr}
		\hline
		Fig.$\sharp$ & 11 & 12 & 13 & 17 & 19 & 20 & 22 & 23 & 25 & 26 \\
		Tab.$\sharp$ & 8 & 9 & 10 & 12 & 13 & 14 & 15 & 16 & 17 & 18 \\
		$s$ & 0.197 & 0.201 & 0.162 & 0.201 & 0.200 & 0.195 & 0.201 & 0.212 & 0.567 & 0.210 \\
		$p_{th}$ & 0.39 & 0.25 & 0.13 & 0.22 & 0.18 & 0.18 & 0.17 & 0.48 & 0.96 & 0.36 \\ 
		$|s-p_{th}|$ & 0.193 & 0.049 & 0.032 & 0.019 & 0.020& 0.015 & 0.031 & 0.268 & 0.393 & 0.150 \\
		Reliability & poor & good & good & good & good & good & good & poor & poor & poor \\ \hline
	\end{tabular}
\caption{Summary of climatological base rate ($s$) and threshold probability maximizing $PSS$ ($p_{th}$) appeared in Barnes \etal (2016). Fig.$\sharp$ and Tab.$\sharp$ are figure numbers and table numbers in Barnes \etal (2016), respectively.}
\label{tbl:flare_model}
\end{table}

\section*{Appendix}
The beta distribution is expressed as 
\begin{equation}
	Be(x;a,b)=\frac{\Gamma(a+b)}{\Gamma(a) \Gamma(b)} x^{a-1} (1-x)^{b-1} \quad (a,b > 0),
\end{equation}
where $\Gamma$ shows a gamma function, and $a$ and $b$ are shape parameters. When the conditional probability density functions $p_1(x)$ and $p_0(x)$ are defined as 
\begin{equation}
	p_1(x)=Be(x;a,b),
\label{eq:p1}
\end{equation}
and
\begin{equation}
	p_0(x)=\frac{1}{1-s}Be(x;a-\beta,b)-\frac{s}{1-s}Be(x;a,b),
\label{eq:p0}
\end{equation}
the theoretical reliability curve $g(x)$ is derived as 
\begin{equation}
	g(x)=\alpha x^{\beta},\quad \alpha:= s \frac{\Gamma(a+b)\Gamma(a-\beta)}{\Gamma(a+b-\beta)\Gamma(a)}.
\label{eq:rel}
\end{equation}
\par

In case (1) mentioned in Section \ref{sec:case1}, the parameters are set as $a=1.1$, $b=0.9$, and $\beta=1$, from which $\alpha=1$ is derived using the equation (\ref{eq:rel}) when $s=0.1$. Therefore, the theoretical reliability curve is $g(x)=x$, which means perfect reliability. $PSS$ is maximum at the threshold probability of 0.1, which is a climatological base rate. \par

In case (2) mentioned in Section \ref{sec:case2}, we set $a=2.2$, $b=0.4$, and $\beta=2$, and $\alpha=0.4$ is derived using the equation (\ref{eq:rel}) when $s=0.1$. These parameters yield the theoretical reliability curve $g(x)=0.4x^2$. The maximum $PSS$ is realized for $g(x)=s$, that is, at a threshold probability of 0.5. \par

In case (3) mentioned in Section \ref{sec:case3}, we set $a=0.83$, $b=1.19$, and $\beta=0.6$, and $\alpha\approx 0.398\approx s^{0.4}$ is derived using the equation (\ref{eq:rel}) when $s=0.1$. In this case, the theoretical reliability curve is $g(x)=s^{0.4}x^{0.6}$. $PSS$ is maximum at $x=s$, that is, a threshold probability of 0.1. \par

\begin{acknowledgements}
I would like to thank Dr. KD. Leka for a valuable comment on the use of the data in Barnes \etal (2016). I also would like to thank two anonymous referees and editor for useful comment on the manuscript. The editor thanks two anonymous referees for their assistance in evaluating this paper. This work was supported partly by MEXT/JSPS KAKENHI Grant Number JP15H05813.
\end{acknowledgements}




\end{document}